\documentclass[twocolumn,aps,pre,amsmath,showpacs]{revtex4}
\usepackage{graphicx}
\usepackage{dcolumn}%
\usepackage{bm}%
\usepackage{amssymb}
\usepackage{amsmath}
\usepackage{color}

\newcommand{\uu}[1]{\underline{\underline{#1}}}
\def\be{\begin{equation}}
\def\ee{\end{equation}}

\begin{document}

\title{Approximate analytical description of the nonaffine response of amorphous solids}
\author{Alessio Zaccone$^{1}$ and Enzo Scossa-Romano$^{2}$}
\affiliation{${}^1$Department of Physics, Cavendish Laboratory, University of Cambridge, JJ Thomson Avenue, Cambridge CB3 0HE, United Kingdom}
\affiliation{${}^2$Department of Chemistry and Applied Biosciences, ETH Zurich, CH-8093 Z\"urich, Switzerland}
\date{\today}
\begin{abstract}
An approximation scheme for model disordered solids is proposed which leads to the fully analytical evaluation of the elastic constants under explicit account of the inhomogeneity (nonaffinity) of the atomic displacements. The theory is in
quantitative agreement with simulations for central-force systems and predicts the vanishing
of the shear modulus at the isostatic point with the linear law $\mu
\sim (z-2d)$, where $z$ is the coordination number. The vanishing
of rigidity at the isostatic point is shown to be the consequence of the canceling out of positive affine and negative nonaffine
terms.

\end{abstract}
 \maketitle

\section{\textbf{Introduction}}
Disordered or amorphous solids represent a
great part of ordinary matter (e.g., glass), including
biological matter (e.g., cytoskeletal networks)
\cite{alexander,zippelius,frey}. Yet, the relationship between
rigidity and disorder has remained elusive and no theory has
hitherto proved able to correctly describe their elastic constants.
The rigidity of disordered solids is also intimately related to the
fundamentally unsolved problem of the glass transition
\cite{reichman}. Elastic rigidity in supercooled liquids emerges from the fluid state at the glass transition without any detectable lowering of the symmetry (apart from translational and replica symmetry-breaking), as opposed to what happens in "ordinary" liquid-solid phase transitions~\cite{anderson}. In recent years it has been recognized that the
intrinsic "softness" of disordered solids is related to the
\emph{nonaffinity} of the atomic displacements
\cite{alexander,reichman,barrat,lubensky}: the atoms in a strained
disordered solid are \emph{not} displaced \emph{proportionally} to
the global strain. The calculation of the contribution to rigidity
due to such nonaffine displacements poses formidable difficulties because
it requires the analytical knowledge of the eigenmodes of the
dynamical or Hessian matrix of the system~\cite{lemaitre,vanhecke},
which is a sparse random matrix. The simplest disordered solids where nonaffinity is supposed to play a significant role are those with nearest-neighbor central-force interactions~\cite{vanhecke,wyart}. In these systems, it is well-known that the shear modulus $\mu$ vanishes at the isostatic point where each particle has an average number of nearest neighbors (mechanical contacts) $z=2d$ \cite{vanhecke,wyart}. Although it seems reasonable from constraint-counting arguments that rigidity is lost when $z=2d$~\cite{phillips}, the linear law $\mu\sim(z-2d)$ observed in simulation studies of completely disordered solids~\cite{ohern} has remained unexplained and the physics behind it represents a long-standing problem \cite{vanhecke,wyart} where the role of nonaffinity is yet unclear. Further, the well-documented inadequacy of affine theories to describe the elasticity and transport properties of amorphous materials calls for an improved theory beyond the affine approximation \cite{makse}.
In this Letter, we propose an
approximation scheme which gives a well-defined
deterministic limit for the nonaffine contributions to the elastic
constants. This leads to a fully analytical description of the elastic constants
which accounts for the microscopic nonaffinity of the atomic displacements.

\section{\textbf{Formalism}}
In the following, Roman indices are used to label
atoms while Greek indices are used to label Cartesian components.
The summation convention over repeated indices holds throughout for
Greek indices. Bold characters denote vectors in $dN$-dimensional
space ($N$=total number of atoms). We closely follow the notation of
Lemaitre and Maloney~\cite{lemaitre} and we start our analysis from the definition of
a Bravais cell for the disordered lattice. The cell is described by
three Bravais vectors and thus by a matrix
$\underline{\underline{h}}$. The potential depends on the particle
position $\underline{r}_i$ and on the shape of the cell which
enforces boundary conditions. Macroscopically imposed deformations
of the cell are described by changes in the Bravais vectors through
a linear map $\underline{\underline{F}}=\uu{h}~\uu{\mathring{h}}^{-1}$ and the relation
$\uu{h}\,=\,\uu F\,\uu{\mathring{h}}$. We denote quantities in the
reference frame, as well as quantities that are measured with
respect to the reference frame, with a circle. Accordingly, the unit
cell in the reference configuration is given through the matrix
$\underline{\underline{\mathring{h}}}$. In the language of continua
$\underline{\underline{F}}$ is the
deformation gradient tensor. 
The position of atom
$i$ after an affine deformation is given by
 \begin{equation}
 r^\alpha_i\,=\,F_{\alpha\beta}\, R^\beta_i
 \label{eqmod1}
 \end{equation}
As a unique exception to the ring notation, we denote with
$\underline{R}_i$ the position of the atoms in the reference cell.
This relation illustrates the definition of affine displacement as
atom $i$ is displaced \emph{proportionally} to the external
deformation. The position of the atom after a strain is denoted by
$r^\alpha_i(\underline{\underline{F}})$ and differs from
Eq.(\ref{eqmod1}) if the nonaffine displacement is not zero. It is
useful to introduce the particle position
$\underline{\mathring{r}}_i$ for an atom which undergoes both affine
and nonaffine displacements, defined by
\begin{equation}
r^\alpha_i(\underline{\underline{F}})\,=\,F_{\alpha\beta}\,
\mathring{ r}^\beta_i(\underline{\underline{F}}). \label{eqmod2}
\end{equation}
Remark also that $\underline{R}_i=\underline{\mathring{r}}_i(0)$.
Hence, for an affine displacement the position vector
$\underline{\mathring{r}}_i$ is kept fixed while the new position is
determined by $\underline{\underline{F}}$. Any additional
(\emph{nonaffine}) displacement is thus parameterized for a given
strain in terms of $\underline{\mathring{ r}}_{i}(\uu F)$. With
these definitions we can express the potential $\mathcal{
U}(\{\underline{r}_i\},\uu F)$ in the coordinates of the reference
frame as $\mathcal{\mathring{U}}(\{\underline{\mathring{r}}_i\},\uu
F)$ which is defined by
\begin{equation}
\mathring{ \mathcal{U}}(\{\underline{\mathring{r}}_i\}, \uu F)=\mathcal{U}(\{\uu F\,\underline{\mathring{r}}_i\}
,\uu F).\label{eqmod3}
\end{equation}

It is convenient to introduce the Cauchy-Green strain tensor $\uu
\eta =\frac{1}{2}(\uu{ F} ^\top \uu F-\uu I)$ to describe the
deformations since the elastic constants are defined in terms of
second derivatives in $\eta$. The deformation is completely
described by this tensor since the total internal energy of the
solid being deformed can be expressed as $\mathcal{U}(\{|\underline{
r}_{ij}|\})$, i.e. as a functional of the set $\{r_{ij}\}$ of
relative distances between atoms
$r_{ij}=|\underline{r}_{ij}|=|\underline{r}_{i}-\underline{r}_{j}|$,
and $\uu\eta$ describes affine transformations in terms of relative
interatomic distances according to $|\uu F
\,\underline{R}_{ij}|^2\,=\,|\underline{R}_{ij}|^2\,+
\,\underline{R}_{ij}^\top\,\uu \eta\,\underline{R}_{ij}$.

A homogeneous strain of the cell $\uu F$ will first bring each atom
to its affine position $r^\alpha_i\,=\,F_{\alpha\beta}\, R^\beta_i$.
In this affine position the total force acting on atom $i$ is in
general not zero since the neighboring atoms may exert a
non-vanishing force-field due to their affine motion. This is
especially true for disordered solids where the nearest neighbors
are placed at random around atom $i$ so that they transmit
unbalanced forces to $i$ (in an ordered lattice the transmitted
forces balance by symmetry such that this effect is often
negligible). It is in response to these virtual forces that the atom
undergoes an additional motion after it has been displaced affinely
such that the energy released in the process reestablishes (local)
mechanical equilibrium. Hence the system under reversible strain
evolves adiabatically along a trajectory $\underline{ \mathbf{
r}}(\uu \eta)$ that minimizes the mechanical energy for a given
strain $\uu \eta$. If we denote by
$\frac{\mathcal{D}}{\mathcal{D}\eta_{\kappa\chi}}$ the derivative
with respect to adiabatic changes of the strain under the constraint
of mechanical equilibrium, one obtains an equation of motion of the
nonaffine displacement by differentiating the force
$f_{i}^{\alpha}=-\frac{\partial \mathcal{U}}{\partial r_{i}^\alpha}$
evaluated in the true position (where it vanishes). In the limit
$\uu \eta \rightarrow 0$ the equation reads
\begin{equation}
\sum_{j}\,H_{ij}^{\alpha\beta}\,\left.\frac{\mathcal{D}\mathring{r}_j^\beta}{\mathcal{D}\eta_{\kappa\chi}}\right|_{\uu
\eta=0} \,=\,\Xi^{\alpha}_{i,\kappa\chi}\label{eqmod4}
\end{equation}
where we have introduced the Hessian $H_{ij}^{ \alpha\beta}$ and the
affine force field $\Xi^{\alpha}_{i,\kappa\chi}$ given
respectively by
\begin{equation}
\begin{split}
&H_{ij}^{ \alpha\beta}=\left.\frac{\partial^2
\mathring{\mathcal{U}}}{\partial \mathring{r}_{i}^\alpha \partial
\mathring{r}_{j}^\beta} \right|_{\uu\eta=0}=\left.\frac{\partial^2
\mathcal{U}}{\partial r_{i}^\alpha \partial
r_{j}^\beta} \right|_{\uu\eta=0}\\
\Xi^{\alpha}_{i,\kappa\chi}&=-\left.\frac{\partial^2
\mathcal{\mathring{U}}}{\partial\mathring{r}_i^\alpha \partial
\eta_{\kappa\chi}}\right|_{\uu\eta=0}=-\sum_{j}\left.\frac{\partial^2
\mathring{\mathcal{U}}}{\partial \mathring{r}_{i}^\alpha \partial
\mathring{r}_{j}^\beta} \right|_{\uu\eta=0}\frac{\partial F_{\beta\beta'}}{
\partial \eta_{\kappa\chi}} R_j^{\beta'}
\end{split}
\label{eqmod5}
\end{equation}

The elastic
constants are defined by
\begin{equation}
C_{\iota\xi\kappa\chi}=\frac{1}{\mathring{V}}\left.\frac{\mathcal
\partial^2 \mathcal{U}}{\partial\eta_{\iota\xi}
\partial\eta_{\kappa\chi}}\right|_{\uu \eta=0}
\label{eqmod6}
\end{equation}
In order to account for the nonaffine relaxation in the calculation of the elastic constants, the derivatives in Eq.(\ref{eqmod6}) have to be taken along a trajectory of (locally) minimum energy.
Following Lemaitre and Maloney~\cite{lemaitre}, we obtain
\begin{equation}
\begin{aligned}
 C_{\iota\xi\kappa\chi}\,&=\,\frac{1}{\mathring{V}} \left[\frac{\mathcal D}{\mathcal D \eta_{\iota\xi}}\,\left( \frac{\partial  \mathcal{\mathring{ U}}}{\partial \eta_{\kappa\chi}}\,+\, \frac{\partial \mathcal{\mathring{ U}} }{\partial \underline{\mathring{r}}_i}\,\frac{\mathcal{D}\underline{\mathring{r}}_i}{\mathcal{D} \eta_{\kappa\chi}}  \right)\right]_{\eta=0}\\
&=\frac{1}{\mathring{V}} \left( \left.\frac{\partial^2 \mathcal U}{\partial\eta_{\iota\xi}\partial \eta_{\kappa\chi}}\right|_{\eta=0}+ \left.\frac{\partial^2 \mathcal{U} }{\partial \underline{\mathring{r}}_i\,\partial\eta_{\iota\xi}}\right|_{\eta=0}\,\left.\frac{\mathcal{D}\underline{\mathring{r}}_i}{\mathcal{D}\eta_{\kappa\chi}}\right|_{\eta=0}\  \right)\\
&=\frac{1}{\mathring{V}}  \left.\frac{\partial^2 \mathcal U}{\partial\eta_{\iota\xi}\partial \eta_{\kappa\chi}}\right|_{\eta=0}\,-\,\frac{1}{\mathring{V}}\,  \underline \Xi_{i,\iota \xi} \left.\frac{\mathcal{D}\underline{\mathring{r}}_i}{\mathcal{D}\eta_{\kappa\chi}}\right|_{\eta=0} \\
&= C_{\iota\xi\kappa\chi}^{A}-C_{\iota\xi\kappa\chi}^{NA}
\end{aligned}
\label{eqmod7}
\end{equation}
where it is evident that the true elastic constant is given by the
affine (Born-Huang) elastic constant $C_{\iota\xi\kappa\chi}^{A}$
corrected by the nonaffine term $-C_{\iota\xi\kappa\chi}^{NA}$. Following Lemaitre and Maloney \cite{lemaitre}, and using Eq. (4) in Eq. (7), one derives the following expression for the nonaffine correction,
\begin{equation}
C_{\iota\xi\kappa\chi}^{NA}=\Xi^{\alpha}_{i,\iota\xi}(H_{ij}^{\alpha\beta})^{-1}\Xi^{\beta}_{j,\kappa\chi}>0
\label{eqmod8}
\end{equation}
The last inequality in Eq. (8) is justified in view of the Hessian matrix being semi-positive definite at mechanical equilibrium. Hence it follows that the correction due to the nonaffine relaxation, $-C_{\iota\xi\kappa\chi}^{NA}<0$, necessarily gives a \emph{negative} contribution to the total rigidity
\cite{lemaitre}.

\section{\textbf{Approximation scheme}}
\subsection{\textbf{The Cauchy bonded-network model}}
Let us consider the disordered Cauchy solid, defined by
the following properties~\cite{alexander}: (\emph{i}) atoms interact pairwise and
only with their nearest neighbors; (\emph{ii}) the interaction
potential is a central-force harmonic potential; (\emph{iii}) the
reference state is \emph{unstressed}, i.e all springs (interatomic
bonds) are relaxed in the minimum of the harmonic well; (\emph{iv})
disorder is spatially decorrelated. The equivalence with a random network of harmonic
springs is evident \cite{vanhecke,vansaarloos}. Hence, the total
free energy is given by $\mathcal U(\{r_{ij}\})= \sum_{\langle
ij\rangle}V_{ij}(r_{ij})$ where the sum runs over all pairs of
nearest-neighbors $\langle ij\rangle$. The pair interaction
potential is given by the harmonic potential
$V(r_{ij})=\frac{\kappa}{2}(r_{ij}-R_0)^2$. $\kappa$ is the atomic
force constant and $R_0$ is the interatomic distance at rest in the
reference frame. Under these conditions the Hessian matrix becomes
\begin{equation}
H_{ij}^{\alpha\beta}= \delta_{ij}\sum_{s}\kappa c_{is} n_{is}^\alpha
n_{is}^\beta -(1-\delta_{ij}) \kappa c_{ij} n_{ij}^\alpha
n_{ij}^\beta\label{eqmod9}
\end{equation}
where we used the identity
$\partial/\partial\underline{r}_{ij}=\underline{n}_{ij}\partial/\partial
r_{ij}$, with $ \underline{n}_{ij}=\underline{r}_{ij}/r_{ij}$.
Further, $c_{ij}$ is the (random) occupancy matrix with $c_{ij}=1$ if
$i$ and $j$ are nearest neighbors and $c_{ij}=0$ otherwise. $c_{ij}$ is a matrix where each row and each column have on average $z$ elements equal to 1 distributed randomly under the constraint that the matrix be symmetric. Using
this form of the Hessian one obtains the affine part of the
elastic constant as
\begin{equation}
C_{\iota  \xi \kappa
\chi}^A=\frac{R_0^2\kappa}{2\mathring{V}}\sum_{ij}
c_{ij}n_{ij}^\iota n_{ij}^\xi n_{ij}^\kappa
n_{ij}^\chi\label{eqmod10}
\end{equation}
which is the well-known Born-Huang formula \cite{alexander,lubensky,lemaitre}.
Further, we also obtain a microscopic expression for the affine
force field from Eq.(\ref{eqmod5}) as $\Xi^{\alpha}_{i,\kappa\chi}
=-\sum_{j} R_{ij} \kappa c_{ij}n_{ij}^\alpha n_{ij}^\kappa
n_{ij}^\chi$. We can now turn to the nonaffine part of the elastic
stiffness, $C_{\iota\xi\kappa\chi}^{NA}$. The Hessian is a
$dN\times dN$ symmetric semi-positive definite matrix with $d$
eigenvalues equal to zero which are due to the global translational invariance of
the solid. Eq.(\ref{eqmod4}) can be solved by normal
mode decomposition which leads to \cite{lemaitre}
\begin{equation}
C_{\iota\xi\kappa\chi}^{NA}=\frac{1}{\mathring{V}}\sum_{\substack{k\\
\lambda_k\neq0}}\,\frac{(\mathbf{\underline{
\Xi}}_{\iota\xi},\underline{\mathbf{v}}_{k})(\mathbf{\underline{
\Xi}}_{\kappa\chi},\underline{\mathbf{v}}_{k}) }{\lambda_k}
\label{eqmod11}
\end{equation}
where $\underline{\mathbf{v}}_k$ are the eigenvectors of the Hessian
(which are orthogonal since the Hessian is symmetric), $\lambda_k$
the corresponding eigenvalues and $(,)$ denotes the normal scalar
product on $\mathbb{R}^{dN}$. In the next section, we shall evaluate
the deterministic limit of Eq.(\ref{eqmod11}), which is a
self-averaging quantity \cite{lemaitre}.

\subsection{\textbf{The approximate Hessian and the affine field projection on its eigenmodes}}
A rigorous derivation requires one to first determine the eigenmodes $\underline{\mathbf{v}}_k$ of the Hessian matrix given by Eq.(9) in order to calculate the projection on them of the affine fields in Eq.(11). Thereafter the average over the disorder is taken to get the thermodynamic limit of $C_{\iota\xi\kappa\chi}^{NA}$. However, the Hessian is a random matrix and both
$\underline{\mathbf{v}}_k$ and $\lambda_k$ depend on the realization
of disorder. Also, being the Hessian sparse, there are no analytical
forms even for its statistical spectral distributions. Nevertheless, the
deterministic limit of Eq.(\ref{eqmod11}) can be calculated
analytically within the following approximation that we propose
here.

Our approximation consists in performing a disorder-average of the orientation-dependent part of the Hessian first and then use the result to calculate the eigenmodes, their inner products with the affine fields and finally the nonaffine correction. Inverting the sequence of "calculating" and "averaging" is sometimes referred to as an effective medium approximation and is not at all unusual in dealing with disordered systems~\cite{batchelor} since it is often the only strategy to keep the treatment analytical. Using some averaged form of the Hessian matrix necessarily implies sacrificing some details of the vibrational spectrum. This problem is addressed in section IV.B and IV.C where we show what details are lost and we study the validity and limitations of the approximation.

In $d=3$ it is
 $\underline{n}_{ij}=(\cos \phi_{ij}\sin
\theta_{ij},\sin \phi_{ij} \sin \theta_{ij},\cos \theta_{ij})$ and the
pair of angles $\phi_{ij}$ and $\theta_{ij}$ univocally specifies
the orientation of the bond $\langle ij\rangle$. The orientation-dependent factors in the Hessian, $n_{ij}^\alpha n_{ij}^\beta$ in Eq. (9), for a large system with uncorrelated isotropic disorder (where every bond can take any orientation in the solid angle with the same probability $1/4\pi$), can be replaced with its isotropic (angular) average, i.e. $n_{ij}^\alpha
n_{ij}^\beta\Rightarrow\delta_{\alpha\beta}/d$. Within this approximation, the Hessian becomes
\begin{equation}
H_{ij}^{\alpha\beta}=\frac{\kappa}{d}\left(\delta_{ij}\sum_j c_{ij}-
(1-\delta_{ij})c_{ij}\right)\delta_{\alpha\beta} \label{eqmod12}
\end{equation}
Remark that this is still a sparse random matrix because of the
positional disorder in the random coefficients $c_{ij}$.
According
to Eq.(\ref{eqmod12}), let us define
$\underline{\underline{H}}=\underline{\underline{\tilde
H}}\otimes\underline{\underline{I}}$ where
$\underline{\underline{I}}$ is the $d\times d$ identity matrix (which represents $\delta_{\alpha\beta}$) and $\underline{\underline{\tilde
H}}$ is the matrix which multiplies $\delta_{\alpha\beta}$ in Eq.(12).
Denoting with $\{\underline{a}_{q}\}_{q=1..N} $ the set of
eigenvectors of $\uu{\tilde{H}}$, which is an orthonormal basis (ONB) of $\mathbb{R}^N$,
and with  $\{\underline{e}_l\}_{l=1..d}$ the  standard Cartesian
basis of $\mathbb{R}^d$, it follows that
$(\underline{\underline{\tilde
H}}\otimes\underline{\underline{I}})(\underline{a}\otimes\underline{e})=\lambda(\underline{a}\otimes\underline{e})$
and thus the $dN$ dimensional set  $\{ \underline{a}_q \,
\underline{e}_l\}_{q=1..N,l=1..d}$ is an ONB  of eigenvectors of
$\uu{H}$ as given by Eq.(\ref{eqmod12}). This allows us to write
(with $\underline{\mathbf{v}}=\underline{a} \, \underline{e}_l$ for
some $\underline{a} \in \{\underline{a}_{q}\}_{q=1..N}$):
\begin{equation}
\begin{aligned}
\left  (\mathbf{\underline{
\Xi}}_{\iota\xi},\underline{\mathbf{v}}\right)\left(\mathbf{\underline{
\Xi}}_{\kappa\chi},\underline{\mathbf{v}}\right)&= \left(\sum_{r}^N
a_r \underline{\Xi}_{r,\iota\xi} \underline{e}_l
\right)\,\left(\sum_{r}^N a_r \underline{\Xi}_{r,\kappa\chi}  \,
\underline{e}_l \right) \\
&=\kappa^2 R_0^2 \,\sum_{r\,s\,r'\,s'}\,\{(a_ra_{r'}\,
c_{rs}c_{r's'})\\
&\times(n_{rs}^l n_{rs}^\iota n_{rs}^\xi n_{r's'}^l n_{r's'}^\kappa
n_{r's'}^\chi)\}
\end{aligned}
\label{eqmod13}
\end{equation}
With our isotropic approximation, we replace the orientation-dependent terms with their
isotropic angular-averaged values which gives $ n_{rs}^l
n_{rs}^\iota n_{rs}^\xi \,n_{r's'}^l n_{r's'}^\kappa
n_{r's'}^\chi=\left(\delta_{rr'}\delta_{ss'}-\delta_{rs'}\delta_{sr'}\right)\cdot
B_{l,\iota\xi\kappa\chi}$ where the $B_{l,\iota\xi\kappa\chi}$ are
geometric coefficients resulting from the angular average. For $d=3$ and $d=2$ they are as follows
\begin{equation}
\begin{tabular}{c||c|c|c|c||c|c|c}
& \multicolumn{4}{c||}{$d=3$}& \multicolumn{3}{c}{$d=2$}\\ \cline{2-8}
$l$		& $x$ 		&	 $y$ 	& 	$z$	&$\sum_{l}$&	$x$	&$y$		&$\sum_l$ \\ \hline
$B_{l,xxxx}$	&$\frac{1}{7}$	& $\frac{1}{35}$&$\frac{1}{35}$ &$\frac{1}{5}$&$\frac{5}{16}$	&$\frac{1}{16}$	&$\frac{3}{8}$\\
$B_{l,xyxy}$	&$\frac{1}{35}$	& $\frac{1}{35}$&$\frac{1}{105}$&$\frac{1}{15}$&$\frac{1}{16}$	&$\frac{1}{16}$	&$\frac{1}{8}$\\
$B_{l,xxyy}$	& $\frac{1}{35}$& $\frac{1}{35}$&$\frac{1}{105}$&$\frac{1}{15}$&$\frac{1}{16}$	&$\frac{1}{16}$	&$\frac{1}{8}$\\
 \end{tabular}
\label{eqmod14}
\end{equation}
Substituting in Eq.(\ref{eqmod11}) we obtain
\begin{equation}
\begin{aligned}
\left  ( \mathbf{\underline{
\Xi}}_{\iota\xi},\underline{\mathbf{v}}\right)\left(\mathbf{\underline{
\Xi}}_{\kappa\chi},\underline{\mathbf{v}}\right) &=\kappa^2
R_0^2\,B_{l,\iota\xi\kappa\chi}\\
& \times\left( \sum_{r\,s} a_r^2\, c_{rs}c_{rs}\,-\,\sum_{r\,s}
a_ra_s\, c_{rs}c_{sr} \right)\\
&=\kappa^2
R_0^2\,B_{l,\iota\xi\kappa\chi}\frac{d}{\kappa} \sum_{rs}^N \,a_r a_s \tilde H_{rs}
\end{aligned}
\label{eqmod15}
\end{equation}
where we used that $c_{rs}^2= c_{rs}c_{sr}= c_{rs}$ and the
identities $\sum_{r}^N a_r^2\sum_{s} c_{rs} -\sum_{rs} a_r a_s
c_{rs}=\sum_{rs}^N a_r a_s [ ( \sum_{j}^N
c_{rj})\delta_{rs}-c_{rs}(1-\delta_{rs}) ]=\frac{d}{\kappa}\sum_{rs}^N a_r a_s \tilde
H_{rs}$. Recalling that $\sum_s^N\tilde H_{rs}\,a_s=\lambda a_r$, we
obtain  $\left  ( \mathbf{\underline{
\Xi}}_{\iota\xi},\underline{\mathbf{v}}_{k}
\right)\left(\mathbf{\underline{
\Xi}}_{\kappa\chi},\underline{\mathbf{v}}_{k}\right) = d\kappa
R_0^2\, \lambda_{k}B_{l,\iota\xi\kappa\chi}$.

\section{\textbf{Results and discussion}}
\subsection{\textbf{Elastic moduli}}
\begin{figure}
\includegraphics[width=0.8\linewidth]{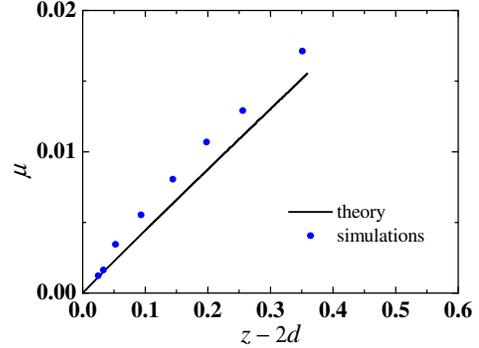}
\caption{(color online). The theoretical predictions for the shear
modulus, Eq. (17), and the simulation results of Ref.\cite{ohern}.
No fitting parameter is used in the comparison. $\kappa=1,
\,R_{0}=1$.} \label{fig1}
\end{figure}

Hence, we have shown that within the isotropic approximation of the Hessian, Eq.(\ref{eqmod12}), the
nonaffine part of the elastic stiffness, Eq.(\ref{eqmod11}), has the
following thermodynamic limit
\begin{equation}
\begin{aligned}
\langle C_{\iota\xi\kappa\chi}^{NA}\rangle &=\frac{1}{\mathring{V}}
\sum_{q=1}^{N}\sum_{l=1}^d\,\frac{ d\kappa R_0^2\,
\lambda_q\,B_{l,\iota\xi\kappa\chi}} {\lambda_q} \\
&=d \frac{N}{\mathring{V}} \kappa R_0^2
\,\sum_{l=1}^d\,B_{l,\iota\xi\kappa\chi}.
\end{aligned}
\label{eqmod16}
\end{equation}
The affine part of the elastic constants for the
disordered Cauchy solid can be obtained by performing the disorder
average of Eq.(\ref{eqmod10}), $\langle
C_{\iota\xi\kappa\chi}^{A}\rangle$, where $\langle .\rangle$ denotes
the angular average, and we always use the isotropic distribution of the bond orientations.
In $d=3$ we thus obtain $\mu^{A}=\langle
C_{xyxy}^{A}\rangle=\frac{1}{30}\frac{N}{V}\kappa z R_{0}^2$, for
the shear modulus, and $K^{A}=\frac{1}{3}(\langle
C_{xxxx}^{A}\rangle+2\langle
C_{xxyy}^{A}\rangle)=\frac{1}{18}\frac{N}{V}\kappa z R_{0}^2$, for
the bulk modulus. Therefore,
using these affine moduli together with the coefficients of Eq.(\ref{eqmod14}) and with Eq.(16)
we derive expressions for the shear and bulk modulus of the
$d=3$ disordered Cauchy solid, respectively as
\begin{equation}
\begin{aligned}
 \mu&= \mu^{A}-\mu^{NA}= \frac{1}{30}\frac{N}{V} \kappa R_0^2 (z-6)\\
 K&=K^{A}-K^{NA}= \frac{1}{18}\frac{N}{V} \kappa R_0^2 (z-6)
\end{aligned}
\label{eqmod17}
\end{equation}
For $d=2$ we obtain
\begin{equation}
\begin{aligned}
 \mu&= \mu^{A}-\mu^{NA}= \frac{1}{16}\frac{N}{V} \kappa R_0^2 (z-4)\\
 K&=K^{A}-K^{NA}= \frac{5}{48}\frac{N}{V} \kappa R_0^2 (z-4)
\end{aligned}
\label{eqmod18}
\end{equation}
Generalizing this result to arbitrary space dimensions
gives the following scaling for the moduli in $d$ dimensions
\begin{equation}
\mu \sim K\sim(z-2d) \label{eqmod19}
\end{equation}
The
predictions of Eq.(\ref{eqmod17}) for the shear modulus, without
fitting parameters, can be compared with the simulations of
Ref.\cite{ohern} of $d=3$ disordered packings of (monodisperse)
compressible spheres interacting via harmonic repulsion in
Fig.(\ref{fig1}).
In the simulations the spheres, at $T=0$, are
slightly compressed to packing fractions $\phi$ just above the so
called jamming point at $\phi_{J}=0.64$ which is also an isostatic
point with $z_{J}=2d$ \cite{ohern,vanhecke}.
Further, the jamming point is a
zero-stress point \cite{ohern}, and the effect of stress on the
global rigidity is therefore small.
However, it seems from this
comparison, and from our results, that stresses, in general, are not
likely to affect the \emph{qualitative} behavior of the
\emph{global} rigidity as they play no role in arriving at the
fundamental scaling law
Eq.(\ref{eqmod18}).

Finally, we note that the vanishing of $K$ at the isostatic point predicted by our theory does not agree with the scaling of $K$ observed in soft-sphere packings where it remains finite at $z_{J}=2d$~\cite{ohern,vansaarloos}. It agrees however with the behavior of random networks where $K$ vanishes linearly at $z_{J}=2d$~\cite{vansaarloos}. The reason for this might be tentatively identified with the fact that in our theory, just like in networks, excluded volume effects are irrelevant, whereas they are important in packings~\cite{vansaarloos}. 

\subsection{\textbf{Vibrational density of states and validity of the approximation}}
The isotropic Hessian matrix introduced for the calculation of the nonaffine contribution to the elastic moduli can be used to obtain the density of vibrational states (DOS) numerically. Recall that the approximate Hessian is given by the following random matrix
\begin{equation*}
H_{ij}^{\alpha\beta}=\frac{\kappa}{3}\left(\delta_{ij}\sum_j c_{ij}-
(1-\delta_{ij})c_{ij}\right)\delta_{\alpha\beta}
\end{equation*}
The Hessian is defined by the  coefficients $c_{ij}$ which depend on the realization $\sigma$ and are, therefore, random variables.
The eigenvalues and thus the eigenvalue distribution of a random matrix are also random quantities.
The explicit calculation of the eigenvalues as functions of the matrix elements is not possible. The approach to the eigenvalue problem in random matrix theory makes use of the self-averaging assumption that the eigenvalue distribution becomes deterministic in the limit of an infinite system size~\cite{lifshitz}. As analytical solutions for the eigenvalue distribution in our case are not possible (due to the sparseness of the Hessian), we resort to a numerical analysis assuming that the self-averaging property holds. Therefore, we can define a limiting eigenvalue distribution $\rho (\lambda)$ as follows
\begin{equation*}
\lim_{N\longrightarrow\infty}\langle\rho^N(\lambda)\rangle_\sigma=\rho(\lambda)
\end{equation*}
Then we have for all realizations $\sigma$ that
\begin{equation*}
\lim_{N\longrightarrow\infty} \,\rho_N(\lambda)[\sigma]=\rho(\lambda)\,.
\end{equation*}
Remark that for a finite $N$ the eigenvalues distribution is discrete and given by
\begin{equation*}
 \rho^N(\lambda)[\sigma]\,=\,\frac{1}{N}\sum_{i=1}^N\,\delta(\lambda-\lambda_i)
\end{equation*}
where $\delta $ is the Dirac delta function. In the limit $N\longrightarrow\infty$, the set of eigenvalues has infinite elements, and the distribution becomes continuous. The factor $1/N$ is necessary  to normalize the density given that the normalization condition is $ \int_0^\infty\,\rho(\lambda)\,d\lambda\,=\,1$.
To analyze numerically the eigenvalue distribution we  calculate the eigenvalues sets $\{\lambda_i\}_r$ for large systems ( $N\simeq10000$ ). Then we create  an  histogram  of the eigenvalues set and we fit it with a continuous curve which approximates the limiting eigenvalue distribution.
The eigenvalue distribution is usually described in terms of the vibrational DOS which we denote as $D(\omega)$, where $\omega$ is the vibrational frequency. The latter is related to the eigenvalue distribution by the change of variables
\begin{equation*}
 \lambda \rightarrow \omega=\sqrt{\frac{\lambda}{m}} \qquad \textrm{and its inverse} \qquad  \omega \rightarrow  \lambda =m\omega^2\\
 \end{equation*}
Hence, with $d\omega=\frac{1}{2\sqrt{m\lambda}} \, d\lambda$ and $ d\lambda =2 m\omega \, d\omega$ we get  that the density distributions of $\lambda$ and $\omega$ are related by:
\begin{equation*}
D(\omega)\,= \rho(m\omega^2)\,2\omega \qquad \textrm{and} \qquad \rho(\lambda)\,=\,\frac{D(\sqrt{\frac{\lambda}{m}})}{2 \sqrt{m\lambda}}
\end{equation*}
\begin{figure}
\includegraphics[width=0.8\linewidth,angle=+90]{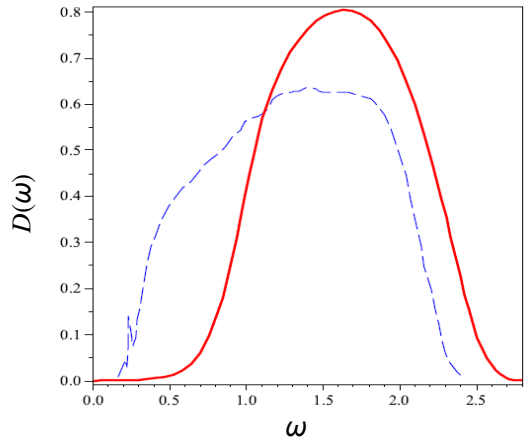}
\caption{(color online). 1: The DOS calculated for a system of $N=10000$ particles according to Eq. (1) (solid line) and $z=8$. 2: Simulations for repulsive unstressed harmonic packings of \cite{xu} with $\phi-\phi_{c}=0.1$ which corresponds to $z-6\simeq2$.} \label{fig2}
\end{figure}
We compare the so obtained DOS from Eq. (12) with the DOS from simulations of unstressed harmonic packings~\cite{xu} where $\phi-\phi_{c}=0.1$ ($\phi_{c}\approx0.64$ is the jamming packing fraction), corresponding to $z-6\simeq2$. The comparison is shown in Fig.2. While the upper end and the main features of the spectrum (width and average) are quite well reproduced, this was somewhat expected because sacrificing information about the bond orientations does not change the single-particle parameters (of order $\sqrt{\kappa/\omega}$) which dominate the highly (Anderson) localized high-$\omega$ modes. On the other hand, it is now clear from this comparison what details of the vibrational spectrum went lost in our isotropic approximation of the Hessian, Eq. (12): the spectrum calculated with the isotropic Hessian is significantly depleted of low-frequency modes whose density is significantly underestimated in comparison with the DOS of the packings. This is probably related to the anisotropic character of the local correlations between particles in the packings (ultimately related to excluded-volume and static effects) which play an important role in the low-frequency modes and that are lost in the approximation. This observation hints again at the possibility that overall our model describes random networks (where excluded-volume are absent) better than sphere packings. The relationship between these observations and the excluded-volume effects should be more systematically investigated in future studies. 
This was already noted in relation to the bulk modulus prediction of our theory which agrees indeed with the random-network scaling ($K\sim z-6$), though not with the sphere packing one, for arguably similar reasons.

In view of these considerations, it is natural to ask why our theory, which seems to underestimate the low-$\omega$ modes, still yields a correct prediction of the shear modulus for sphere packings. This question is related to the issue of the role played by the low-frequency modes in the nonaffine response. While this issue is a very open and unsolved one in our current understanding of amorphous solids~\cite{wyart}, a tentative, and certainly incomplete, answer to this deep question is proposed in the next section.

\subsection{\textbf{On the role of low-frequency modes in the nonaffine response}}
To assess the relative importance of different regimes of the vibrational spectrum in the nonaffine elastic response, it is instructive to rewrite the elastic moduli with the nonaffine correction in the continuous frequency domain. According to the nonaffine linear response formalism of Lemaitre and Maloney~Ref.~\cite{lemaitre}, in the thermodynamic limit one has
\begin{equation}
\langle C_{\iota\xi\kappa\chi}\rangle=\langle C_{\iota\xi\kappa\chi}^{A}\rangle-\int_{0}^{\infty}\frac{D(\omega)\Gamma_{\iota\xi\kappa\chi}(\omega)}{m\omega^{2}}~d\omega
\label{eqmod20}
\end{equation}
where the correlators on the frequency shells are defined by
\begin{equation}
\Gamma_{\iota\xi\kappa\chi}(\omega)=\langle(\mathbf{\underline{
\Xi}}_{\iota\xi},\underline{\mathbf{v}}_{k})(\mathbf{\underline{
\Xi}}_{\kappa\chi},\underline{\mathbf{v}}_{k})\rangle_{\omega_{k}\epsilon[\omega,\omega+d\omega]}
\label{eqmod21}
\end{equation}
The function $\Gamma_{\iota\xi\kappa\chi}(\omega)$ thus represents the projection of the affine fields on the frequency shells and its magnitude gives the importance of the contribution of each frequency shell to the nonaffine response. From Eq.(20) it is evident first of all that in the zero-frequency limit $\omega\rightarrow0$ the moduli diverge to minus infinity, i.e. $\langle C_{\iota\xi\kappa\chi}\rangle\rightarrow-\infty$, unless either $D(\omega=0)=0$ or $\Gamma_{\iota\xi\kappa\chi}(\omega)=0$. At the isostatic or jamming point of sphere packings one has that the DOS develops soft modes with $D(\omega=0)\neq0$. As the nonaffine linear formalism is an exact theory, it is then strictly necessary that
\begin{equation}
\lim_{\omega\rightarrow0}\Gamma_{\iota\xi\kappa\chi}(\omega)=0
\label{eqmod22}
\end{equation}
i.e. the zero-frequency modes must not contribute to the nonaffine response. This is what has been observed indeed in the numerical simulations of Ref.~\cite{lemaitre} where, in the case of a Lennard-Jones glass, the function $\Gamma_{\iota\xi\kappa\chi}(\omega)$ measured in the simulations goes to zero at $\omega=0$. Furthermore, in the same simulation study~\cite{lemaitre}, it was found that $\Gamma_{\iota\xi\kappa\chi}(\omega)$ not only goes to zero at zero frequency, but is a monotonically growing function of $\omega$ in the entire domain, such that it has significantly lower values at low $\omega$ than in the middle and upper part of the spectrum where it reaches its maximum value (cfr. Fig.5 in Ref.\cite{lemaitre}). Hence, the simulation results of ~\cite{lemaitre} indicate that the contribution of low-frequency modes to the nonaffine response is small whereas the leading contribution comes from the high-frequency modes. This observation is also in agreement with physical intuition: the source of the nonaffine response is given by the projection of the affine fields on the eigenmodes which has a higher value the more energetic the modes are.

Based on these observations, one can conclude that the low-frequency modes play a relatively minor role in the nonaffine response as compared to the high-frequency modes. This explains why our theory, which underestimates the low-frequency modes in the case of sphere packings, still yields correct predictions for the shear modulus in excellent agreement with simulations (Fig.1).

In the case of the bulk modulus, simulations~\cite{lemaitre} give practically the same behavior for the correlator $\Gamma_{\iota\xi\kappa\chi}(\omega)$ as for the shear modulus, with the low-frequency modes contributing to the nonaffine response to a minor extent. In this case, the failure of the theory in predicting the correct scaling for sphere packings (despite being successful for networks) is more likely to be ascribed to the geometric attenuation of the random affine fields under hydrostatic pressure due to excluded volume, as we speculated in section IV.A. However, this hypothesis has to be tested in future work by means of ad hoc numerical studies as the bulk modulus scaling of packings is a problem currently under debate~\cite{vansaarloos}.

\section{\textbf{Conclusion}}
We have developed an approximate, fully analytical theory of the nonaffine elastic response of amorphous solids which explicitly takes into account the nonaffinity of the atomic displacements. We have applied the nonaffine linear formalism in the formulation of Lemaitre and Maloney~\cite{lemaitre} to the so-called Cauchy bonded-network model~\cite{alexander}, i.e. to networks of harmonic central-force springs. In order to evaluate the nonaffine correction to the elastic moduli analytically, an approximation of the Hessian matrix has been proposed where the bond orientation-dependent factors in the Hessian are replaced with their isotropic average (isotropic Hessian). Even though the isotropic Hessian has a density of states which significantly lacks low-frequency modes in comparison with sphere packings, our approximation yields predictions of the shear modulus in excellent quantitative agreement with simulations of sphere packings~\cite{ohern}. The good agreement is explained with the observation, supported by simulations in the literature~\cite{lemaitre}, that the low-frequency modes, underestimated by our approximation, play a relatively minor role in the nonaffine response, which is controlled by the upper part of the vibrational spectrum (that is well reproduced by our theory).
While our approximation is not suited to accurately describe transport properties of disordered solids in the low-connectivity and low-frequency limits~\cite{vitelli}, it seems on the other hand successful in accurately describing the elastic response to shear of disordered solids. Furthermore, our theory provides a completely new insight into the linear vanishing of shear rigidity at the isostatic point ($z=2d$) of disordered solids: this happens because the nonaffine correction at the isostatic point becomes
equal in absolute value, but with opposite sign, to the affine part of the
shear modulus.

{\it Aknowledgements.}
A.Z. acknowledges financial support by the Swiss National Science Foundation (project no. $PBEZP2-131153$).
Enlightening discussions with V. Vitelli, M. Warner, E.~M. Terentjev, and R. Blumenfeld are gratefully acknowledged.
We are very thankful to M. Morbidelli for generously supporting this project in the initial phase.


\begin{thebibliography}{99}

\bibitem{alexander} S. Alexander, Phys. Rep. {\bf 296}, 65 (1998).
\bibitem{zippelius} P. M. Goldbart, H. E. Castillo, and A.
Zippelius, Adv. Phys. {\bf 45}, 393 (1996).
\bibitem{frey} C. Heussinger and E. Frey, Phys. Rev. Lett. {\bf 96}, 017802 (2006).
\bibitem{reichman} A. Widmer-Cooper et al., Nat. Phys. {\bf 4}, 711
(2008); E. Del Gado et al., Phys. Rev. Lett. {\bf 101}, 095501
(2008).
\bibitem{anderson} P.W. Anderson, in \emph{Ill-Condensed Matter} Les Houches Session XXXI, Eds. R. Balian, R. Maynard, G. Toulouse (North-Holland, Amsterdam, 1979).
\bibitem{barrat} A. Tanguy, et al. Phys. Rev. B {\bf 66}, 174205 (2002).
\bibitem{lubensky} B. A. DiDonna and T. C. Lubensky, Phys. Rev. E
{\bf 72}, 066619 (2005).
\bibitem{lemaitre} A. Lemaitre and C. Maloney, J. Stat. Phys. {\bf
123}, 415 (2006).
\bibitem{vanhecke} M. van Hecke, J. Phys.: Condens. Matter {\bf 22},
033101 (2010).
\bibitem{wyart} M. Wyart, Ann. Phys. (Paris) {\bf 30}, 1 (2005); W. Ellenbroek et al., Phys. Rev. Lett. {\bf 97}, 258001 (2006).
\bibitem{phillips} J.C. Phillips and M.F. Thorpe, Sol. State Commun. \textbf{53}, 699 (1985).
\bibitem{ohern} C. S. O'Hern et al., Phys. Rev. E {\bf 68}, 011306
(2003).
\bibitem{makse} H. A. Makse et al., Phys. Rev. Lett. {\bf83}, 5070 (1999); H.A. Makse et al., Phys. Rev. E {\bf 70}, 061302 (2004).
\bibitem{vansaarloos} W. G. Ellenbroek et al., EPL {\bf 87}, 34004 (2009).
\bibitem{batchelor} J.C. Phillips, in \emph{Rigidity Theory and Applications} p.155, Eds. M.F. Thorpe and P.M. Duxbury (Kluwer Academic, New York, 1999); L.V. Kantorovich, \emph{Quantum Theory of the Solid State: An Introduction}, p. 257-259 (Kluwer Academic, Dordrecht, 2004); G. K. Batchelor and R. W. O'Brien, Proc. R. Soc. London, Ser. A {\bf355}, 313 (1977).
\bibitem{lifshitz} I. M. Lifshitz, S. A. Gredeskul, and L. A. Pastur, \emph{Introduction to the theory of disordered systems} (New York, Wiley, 1988).
\bibitem{xu} N. Xu, V. Vitelli, A. J. Liu, and S. Nagel, EPL 90, 56001 (2010).
\bibitem{vitelli} N. Xu et al., Phys. Rev. Lett. {\bf 102}, 038001
(2009); V. Vitelli et al., Phys. Rev. E {\bf 81}, 021301 (2010).



\end{thebibliography}
\end{document}